\documentstyle[preprint,aps,eqsecnum]{revtex}

\newcommand{\beq}{\begin{equation}}
\newcommand{\eeq}{\end{equation}}
\newcommand{\beqa}{\begin{eqnarray}}
\newcommand{\eeqa}{\end{eqnarray}}
\newcommand{\dg}{\mbox{$\Delta \Gamma$}}
\newcommand{\adg}{\mbox{$\Delta \Gamma$}}
\newcommand{\dm}{\mbox{$\Delta M$}}
\newcommand{\sign}{{\rm sign}}

\def\npb#1{Nucl.\ Phys.\ {\bf B #1}}
\def\plb#1{Phys.\ Lett.\ {\bf B #1}}
\def\prd#1{Phys.\ Rev.\ {\bf D #1}}

\def\ijmpa#1{Int.\ J.\ Mod.\ Phys.\ {\bf A #1}}

\begin{document}

\draft

{\tighten
\preprint{\vbox{\hbox{WIS-96/13/Mar-PH}
                \hbox{hep-ph/9603244} }}

\title{The $B_s$ width difference beyond the Standard Model}

\author{Yuval Grossman}

\address{ \vbox{\vskip 0.truecm}
  Department of Particle Physics \\
  Weizmann Institute of Science, Rehovot 76100, Israel }

\footnotetext{\footnotesize E-mail: ftyuval@wicc.weizmann.ac.il}

\maketitle

\begin{abstract}%
The Standard Model predicts a large width difference in the
$B_s$ system. 
New physics can contribute significantly to 
the {\it mass} difference. If this contribution 
is CP violating, this will always result in a reduction of
the {\it width} difference.
The analyses of measurements of the width difference 
using $B_s$ decays into CP eigenstates have to be modified
in the presence of new physics.
We discuss how the width difference and the
new CP violating phase in $B_s$ mixing can be measured.

\end{abstract}

} 

\newpage


\section{Introduction}
The physics of the $B_s$ system is entering an exciting era, when
experiments start exploring the range of mixing that is relevant
to the Standard Model (SM) or to models of New Physics (NP).
Recently, Dunietz \cite{Dun} has studied the possibility of	
using untagged $B_s$ samples for various measurements.
The important ingredient in such an analysis
is the large
width difference that is expected in the $B_s$ system.
In some sense, the $B_s$ system is similar to the neutral
kaon system, where experiments study {\it mass} eigenstates
($K_L$ and $K_S$), rather than the $B_d$ system,
where experiments study the {\it flavor} 
eigenstates ($B^0$ and $\bar B^0$).

In \cite{Dun}, several ideas were put forward on how to measure
the width
difference, \dg, and how to extract the CP violating phase $\gamma$.
However, the entire analysis was carried out in the framework of
the SM.
In this paper we investigate the implications of NP which
can significantly affect
the $B_s$ mass difference, \dm. 
Phenomenological constraints on the relevant NP
are rather weak. Moreover, various theories of flavor physics
suggest that NP effects are more likely to appear in processes
involving the heavy generation; the $B_s$ system is unique in
that it is the only neutral meson that is expected to exhibit
mixing and does not involve first generation quarks.

We focus our attention on NP that contributes only to \dm.
This is the case in most extensions of the SM: new, heavy
particles cannot compete with the $W$-mediated tree level
$B_s$ decay amplitudes, and thus to a very good approximation
all decay amplitudes are given by the SM. However, the new effects 
could even dominate over the highly suppressed box diagram 
mixing amplitude. Naively, one would think that since in such models
the $B_s$ decays are described by 
the SM, also the width difference is given by the SM.
As we will discuss in this paper, this is incorrect. 

The organization of the paper is as follows: 
in Section II we collect the relevant formalism. 
In Section III we discuss how
new sources of CP violation
that in general arise in NP models reduce \dg.
In Section IV we describe how the width difference can be determined 
in a model independent way, and we analyze  
several  methods to extract the 
new CP violating phase using untagged $B_s$ samples. 
We also argue that the unitarity angle $\gamma$ can be extracted
even in the presence of NP.  
Finally, Section V contains our conclusions.


\section{Formalism}
We start by collecting the relevant formulae and definitions. Most of
them are well known \cite{Yossi}.
An arbitrary neutral $B_s$-meson state,
\begin{equation}
a \mid B_s\rangle + b\mid \overline{B}_s\rangle\,,
\end{equation}
is governed by the time-dependent Schr\"odinger equation
\begin{equation}
i \;\frac{d}{dt} \;\left(\begin{array}{c} a \\ b \end{array}
\right) = {\bf H}
\left(\begin{array}{c} a \\ b \end{array}
\right) \equiv \left({\bf M} -\frac{i}{2} {\bf \Gamma}\right) \;
\left(\begin{array}{c} a \\ b \end{array}\right) \,.
\end{equation}
Here ${\bf M}$ and ${\bf \Gamma}$ are $2\times 2$ 
Hermitian matrices. 
CPT invariance guarantees $H_{11} = H_{22}$.
The eigenstates of the mass matrix are
\begin{equation}
\mid B_L\rangle=p\mid B_s\rangle + \,q\mid \overline{B}_s\rangle \,, \qquad
\mid B_H\rangle=p\mid B_s\rangle - \,q\mid\overline{B}_s\rangle \,,
\end{equation}
with eigenvalues 
\begin{equation}
\mu_{L,H} =M_{L,H} -\frac{i}{2} \;\Gamma_{L,H} \,.
\end{equation}
Here $M_{L,H}$ and $\Gamma_{L,H}$ denote the masses and decay 
widths of $B_{L,H}$ and, by definition, $M_H>M_L$.
In the SM $\Gamma_{L} > \Gamma_{H}$ and then $B_{L}$
is the short lived mass eigenstate. However, in the presence of
NP this is not necessarily true. Therefore,   
we also label the mass eigenstates by
$B_{long}$ and $B_{short}$, where, 
by definition, $\Gamma_{short}>\Gamma_{long}$. We define
\beq
\dm \equiv M_H-M_L\,, \qquad 
\dg \equiv \Gamma_{short}-\Gamma_{long}\,,\qquad
\Gamma \equiv\frac{\Gamma_L +\Gamma_H}{2} \,.
\eeq
The coefficient ratio $q/p$ satisfies
\begin{equation}
\left(\frac{q}{p}\right)^2 =
\frac{(M_{12}^* -\frac{i}{2} \Gamma_{12}^*)}
 {(M_{12} -\frac{i}{2} \Gamma_{12})} \,.
\end{equation}
The width difference is 
\begin{equation} \label{mastf}
\Delta\Gamma =\frac{4 |\,Re \,(M_{12} \Gamma^*_{12})|}{\dm} \,.
\end{equation}
The experimental lower bound
$\dm/\,\Gamma > 8.8$  \cite{expbound}, implies $\dm \gg \dg$ and
consequently \cite{Yossi}
$| M_{12} | \,\gg \,|\Gamma_{12} |$.
Thus, to a very good approximation \cite{Yossi}
\begin{equation} \label{dmrt}
\dm = 2 \left| M_{12} \right| \,, \qquad 
\left|{q/p}\right|=1\,.
\end{equation}
For a final state $f$ 
we define the interference terms
\begin{equation} \label{deflam}
\lambda \equiv \frac{q}{p}\,\frac{ \langle f\mid\overline{B}_s
\rangle }{ \langle f\mid B_s \rangle}, \qquad
\overline{\lambda} \equiv \frac{p}{q} \,\frac{ \langle \overline f \mid
B_s \rangle}{ \langle \overline f \mid\overline{B}_s  \rangle }.
\end{equation}
We further assume $|\langle f\mid B_s \rangle|=
|\langle \overline f \mid\overline{B}_s \rangle|$
(namely, no CP violation in decay) and then 
$|\lambda|=|\overline{\lambda}|$.
We always choose $f$ such that $|\lambda| \le 1$.
The rapid time-dependent oscillations, which
depend on $\dm t$, cancel in untagged data samples \cite{Dun}.
We define
\begin{equation}
\Gamma\left[f\left(t\right)\right]\equiv\Gamma\Big(B_{phys}\left(t\right)
\rightarrow f\Big)+\Gamma\Big(\overline{B}_{phys}\left(t\Big)\rightarrow
f\right) \,,
\end{equation}
where $B_{phys}\; (\overline{B}_{phys}$) denotes
a time-evolved
initially unmixed $B_s \; (\overline{B}_s)$. We have
\begin{eqnarray} \label{genrate}
\Gamma \left[f\left(t\right)\right] & = &
\frac{\Gamma (B_s \rightarrow
f)}{2} \;\bigg\{\left(1 + |\lambda |^2 \right) \;\left(e^{-\Gamma_L t}
+e^{-\Gamma_H t}\right) + 
 2 Re \,\lambda \;\left(e^{-\Gamma_L t} -e^{-\Gamma_H t}
\right)\bigg\}\,, \\
\Gamma \left[\overline f\left(t\right)\right] & = &  
\frac{\Gamma (B_s \rightarrow
f)}{2} \;\bigg\{\left(1 + |\lambda |^2 \right)
\;\left(e^{-\Gamma_L t}
+e^{-\Gamma_H t}\right) + 
2 Re \,\overline \lambda \;\left(e^{-\Gamma_L t} -e^{-\Gamma_H t}
\right)\bigg\} \,. \nonumber
\end{eqnarray}
where we used $|\lambda|=|\overline{\lambda}|$ and
$\Gamma (B_s \rightarrow f)=\Gamma (\overline B_s \rightarrow \overline f)$.
Two limits are of particular interest.
For decay modes which are specific of $B_s$ or $\overline B_s$, 
$\lambda =\overline{\lambda} =0$
(semileptonic decays are
flavor specific; $b\to c\bar ud$ decays are also flavor specific to a
very good approximation),
\begin{equation} \label{slrate}
\Gamma\left[f\left(t\right)\right]=
\Gamma\left[\overline f\left(t\right)\right] =
\frac{\Gamma\left(B_s \rightarrow f\right)}{2} \;
\bigg\{e^{-\Gamma_L t} +e^{-\Gamma_H t} \bigg\} \;.
\end{equation}
For a final state that is a CP eigenstate, 
$\lambda^*=\overline{\lambda}$ with $|\lambda|=1$, and
\begin{equation} \label{cprate}
\Gamma\left[f\left(t\right)\right]  =\frac{\Gamma\left(B_s \rightarrow
f\right)}{2}\bigg\{
e^{-\Gamma_L t} + e^{-\Gamma_H t} +
Re\lambda \left(e^{-\Gamma_L t} -e^{-\Gamma_H t}\right)\bigg\} \;.
\end{equation}


\section{\dg\ in the SM and beyond}
\dg\ is produced by decay channels that are
common to $B_s$ and $\overline{B}_s$.
In the SM, \adg\ is relatively large since
the CKM unsuppressed parton decay $b \to c \bar c s$
produces such final states. 
Two approaches have been used to actually estimate \dg.
The first is a quark level calculation, where \dg\ corresponds
to the imaginary part of the box diagram. 
Assuming factorization \cite{hag}, 
and including QCD corrections \cite{vuks}, the result is
\beq \label{sminc}
{\dg \over \Gamma} \approx 0.20 \left({f_{B_s} \over 210\,{\rm MeV}}
\right)^2.
\eeq
In deriving (\ref{sminc}),
$\dg = 2 \Gamma_{12}$ has been used, which is justified
when CP violation is neglected. This is a very good
approximation in the SM.
The second approach sums over exclusive channels \cite{alek}.
Assuming factorization and Heavy Quark Symmetry relations between form factors,
\beq \label{smexc}
{\dg \over \Gamma} \approx 0.15\,.
\eeq
The explicit dependence on the decay constants was not given in  \cite{alek}.
In general the width difference scales like the decay constant squared. 
Therefore, the use of the  recently measured $f_{D_s} = 273 \pm 39 \,$MeV  
\cite{fdsexp}, instead of $f_{D_s} = 230\,$MeV as used in \cite{alek} 
increases the estimate of \adg.

When NP contributes comparably to or dominates over the
SM contribution to the $B_s$ mixing, CP may be significantly
violated. We now show that new CP violating contributions
to the mixing
always {\it reduce} \adg\ relative to the SM prediction.
Eqs. (\ref{mastf}) and (\ref{dmrt}) lead to
\beq \label{mast}
\Delta \Gamma
=  2 |\Gamma_{12}|\; |\cos2\xi|\,, \qquad
2\xi \equiv \arg(-M_{12} \Gamma^*_{12})\,.
\eeq
Under the reasonable assumption that NP does not significantly
affect the leading decay processes,
$\Gamma_{12}$ arises from  $\Gamma (b \to c \bar c s)$. Consequently,
$2\xi$ is the phase difference between
the total mixing amplitude and the $b \to c \bar c s$ decay  amplitude.
In the SM,
\beq
\xi=\beta^\prime\equiv\arg\left(-{V_{cs}V_{cb}^*\over
V_{ts}V_{tb}^*}\right)\approx 0,
\eeq
and then $\cos2\xi = 1$ to a very high accuracy.
With NP, new phases could be present, leading to $\cos2\xi < 1$.
This proves our statement.

The reduction of \adg\ can be understood intuitively as follows.
In the absence of CP violation, the two mass eigenstates are also
CP eigenstates. 
The large \adg\ is an indication that
most of the $b \to c \bar c s$ decays are into CP even final states.
With CP violation, in the basis where the $b\to c\bar cs$ amplitude
is real, the
mass eigenstates are no longer approximate CP eigenstates. Then,
{\it both} mass eigenstates decay into
CP even final states. Consequently, \adg\ is reduced.

The NP effects on the mixing amplitude $M_{12}$ can be parameterized as
\beq
M_{12} = M_{12}^{SM} \left(1+a e^{i \theta} \right),
\eeq
so that $\xi$ is modified to (we approximate $\beta^\prime=0$)
\beq
2\xi=\arg\left(1+a e^{i \theta}\right)= 
\tan^{-1}\left({a \sin\theta \over 1+ a \cos\theta}\right).
\eeq
The parameters $a$ and $\theta$ give the relative magnitude and
relative phase of the NP contribution, i.e. $a=|M_{12}^{NP}/
M_{12}^{SM}|$ and $\theta=\arg\left(M_{12}^{NP}/M_{12}^{SM}
\right)$.
Phenomenological constraints do not exclude $a>1$ \cite{yossibs}, 
and $\theta$ can have any value. Consequently, in the presence of NP,
$\xi$ could assume any value.

We do not study any particular NP model in detail, but 
mention here a few cases which predict (or can accommodate)
large $a$ and $\theta$.
In models with vectorlike 
down type quarks, a tree level $bsZ$ vertex is induced.
The experimental bounds 
on $BR(B \to X_s \mu^+ \mu^-)$ \cite{PDG} and $BR(B \to X_s \nu \bar \nu)$
\cite{gln} allow for $a \lesssim 0.25$ and arbitrary $\theta$ \cite{sil}.
In fourth generation models the $t'$ box diagram can be 
large and $a>1$ with arbitrary $\theta$  is allowed \cite{yossibs,dln}.
In SUSY models without $R$ parity, tree level 
sneutrino exchange (induced by the $\Delta L=1$ couplings)
contributes to $B_s$ mixing.
Then, $a>1$ with arbitrary $\theta$ is allowed \cite{bgnn}.

Finally, we mention the possibility of NP that affects
the decay rates. An example would be models that give large
$BR(b\to sg) = O(10)\%$ \cite{Kagan}. Since $b \to s g$ can produce
CP eigenstates, $\Gamma_{12}$ could change as well.
In particular, $\dg$ may become somewhat
larger than the SM prediction.
For the rest of the paper, we restrict ourselves to the 
more reasonable case, namely when only $M_{12}$ is 
significantly modified by the NP.

To conclude this section: If NP contributes significantly to the
$B_s$ mixing, the likely results are the enhancement of
\dm\ and a suppression of \adg.
If experiments are able to push the lower bound on \dm\ up, or
the upper bound on \adg\ down, they might, in principle, 
be able to find NP.


\section{Measuring \adg, $\xi$ and $\gamma$}
In this section, we 
examine how to measure \adg\ and the 
angles $\xi$ and $\gamma$ using untagged $B_s$ data samples.
In Ref. \cite{Dun}, several methods of how to measure \dg\ and 
$\gamma$ have been
analyzed in the framework of the SM. Here, 
we study how they have to be modified in the 
presence of NP and how they can probe it.
We start by looking to the methods that were proposed 
to extract \dg. We show that by combining few measurements,
both \adg\ and $\xi$ can be extracted.

The first method is to fit the time dependent rate of flavor specific
decays (see Eq. (\ref{slrate})) to two exponentials. This
fit determines $\Gamma_{long}$ and $\Gamma_{short}$, 
and therefore, $\Gamma$ and \dg.
This method is still valid in the presence of NP.

The second class of methods to measure 
\dg\ uses decays into CP eigenstates.
(Unless otherwise specified, when we talk about CP eigenstates
we refer to those that are produced by the 
$b \to c \bar c s$ decay.)  
For concreteness, we focus on one method. 
We assume that $\Gamma$ is known from either
a fit to a flavor specific data samples, or from 
the average $b$ hadrons lifetime \cite{Dun}.
Then,  in the SM, the $B_s$ lifetime measured using $B_s$ decays
into CP even   final states determines $\Gamma_{short}$, and therefore, \dg.
(Examples of such final states are $D_s^+ D_s^-$ and the CP even 
component of $J/\psi \phi$ that can be obtained using transversity analysis
\cite{Lipkin}.)
As mentioned in \cite{Dun},
these methods are modified in the presence of NP.
We now study this point in detail.

In a general NP model, when decays into 
CP eigenstates are dominated by the $W$ exchange
tree diagram, we have
\beq \label{lamdefxi}
\lambda=\pm e^{2i\xi},
\eeq
where, by convention, $+(-)$ stand for a CP even (odd) final state.
We emphasize that the phase $\xi$ in (\ref{lamdefxi}) and
(\ref{mast}) is the same phase. Then the decay rate into CP 
eigenstates (\ref{cprate}) satisfies
\beq \label{method}
\Gamma(B \to {\rm CP\ even},t) \propto
\cos^2\!\xi \; e^{-\Gamma_L t} +
\sin^2\!\xi \; e^{-\Gamma_H t}.
\eeq
For a decay into a CP odd state, 
$\Gamma_L$ and $\Gamma_H$ are interchanged.

A comment about discrete ambiguities is in order.
In the SM the light state has a shorter lifetime and it is
a CP even state \cite{Dun}.
However, in the presence of NP things may be different.
When the NP contribution is large and negative 
($\cos 2\xi<0$)  
the heavy state has a shorter lifetime, thus
reversing the SM prediction.
For $\sin 2\xi \ne 0$ the mass eigenstates are no longer CP eigenstates,
both eigenstates decay into CP even and CP odd states 
(see Eq. (\ref{method})).
We emphasize, however, that always the longer lived state is
``closer'' to the CP odd state, namely, its decay width into CP odd states 
is larger than the decay width of
the other mass eigenstate. 
Therefore, because $\dm/M$ is tiny,
$\sign(\Gamma_L-\Gamma_H)=\sign(\cos 2\xi)$ is practically undetectable. 
Moreover,
because $2\xi$ always appears as an argument of a cosine function,
also $\sign(\sin 2\xi)$ cannot be determined using untagged $B_s$ samples.
(Note, however, that if the rapid $\dm t$ time dependent CP asymmetries 
in $B_s$ decays will be measured, this sign will be determined,
since asymmetries in modes 
governed by $b \to c \bar c s$ can be used to 
extract $\sin 2\xi$ \cite{Yossi}.)
In conclusion, in our analysis, there is a four fold ambiguity in the value
of $2\xi$.


\subsection{Combining flavor specific and CP eigenstates data}
In principle, a three parameter fit of a decay into a
CP even eigenstate can be used to measure $\Gamma$, \adg\ and $\xi$
using Eq. (\ref{method}). 
Even if this cannot be done in practice, by comparing the measurements 
of \adg\
from flavor specific decays and CP eigenstate decays, $\xi$ can be measured.
Experimentally, most of the data are expected to be taken for small
$\Gamma\, t$. Then, using $\adg \, t \ll 1$, Eq. (\ref{method}) becomes
\beq \label{sltimp}
\Gamma(B \to {\rm CP\ even},t) \propto
e^{-\Gamma_+ \,t}\,, \qquad
\Gamma_+ \equiv
\left(\Gamma + {\dg\,|\cos 2\xi| \over 2} \right)\,.
\eeq
Using $\Gamma$ and \adg\ as measured from the flavor specific data,
a one parameter fit to the decay rate gives $\xi$.
Actually, such a fit 
determines
\beq
\dg_{CP} \equiv 2(\Gamma_+ - \Gamma) = \dg\, |\cos2\xi|\,.
\eeq
By comparing it to the real width difference as measured
using flavor specific data,  $\dg_{FS}$, we get
\beq \label{combi}
|\cos2\xi| = {\dg_{CP} \over \dg_{FS}}\,. 
\eeq
This method would be particularly useful if $\xi$ is neither very
small nor very large. For $\xi \sim \pi/4$ the width difference becomes too
small to be measured (see Eq. (\ref{mast})). 
For $\xi \sim 0$ the precision of the
measurement should be very high.
While we concentrate on one example, we emphasize that Eq. (\ref{combi})
is relevant to all the methods for measuring
\dg\ using decays into CP eigenstates as suggested in \cite{Dun}.


\subsection{Theory}
Since $\xi=0$ in the SM, we get
from Eq. (\ref{mast})
\beq \label{slpr}
\dg=\dg_{SM}  |\cos2\xi|\,.
\eeq
We learn that, if we knew
the SM prediction for $\dg_{SM}$,
the measurement of \dg\ would allow for the determination of $\xi$.
The problem is, of course, that
we do not know  how to calculate $\dg_{SM}$ to a high accuracy.
The box diagram calculation suffers from the uncertainty in $f_{B_s}$ and,
in any case, can be used only as an order of magnitude estimate
since hadronic physics alters it. Similarly, there are large
uncertainties in using the sum over exclusive states.
However, we are able to get bounds on $\xi$ even with these large
uncertainties. To do that, we need only rough bounds on the SM range.
The upper bound is rather safe, as $\dg\le\Gamma(b\to c \bar cs)$.
Based on the inclusive \cite{ball} and the exclusive \cite{bdh}
calculations, 
we can conservatively take an upper bound of $BR(b \to c \bar c s)<40\%$.
The lower bound is less reliable. Even if one can
obtain a lower bound on $BR(b \to c \bar c s)$ it would not help since
we only know that $\dg/\Gamma \le BR(b \to c \bar c s)$.
However,
from Eqs. (\ref{sminc}) and (\ref{smexc}) we can conservatively
take $\dg /\Gamma > 10\%$. We believe then that it is safe to assume
\beq
10\% < {\dg_{SM} \over \Gamma} < 40\%\,.
\eeq
Therefore, a measurement of \dg\ will 
provided an upper bound on $\xi$:
\beq
\cos2\xi > \left({\dg \over \Gamma}\right)_{\rm exp}
{1\over 40\%}\,,
\eeq
and similarly,
any upper bound on \dg\ below the 
SM prediction, would give a lower bound on $\xi$:
\beq
\cos2\xi < \left({\dg \over \Gamma}\right)_{\rm exp}
{1\over 10\%}\,.
\eeq
Note that $\dg/\Gamma<0.1$  would be evidence for NP;
this is an unusual situation. Usually, NP enhances observables.
In this case the SM is at the maximum and NP can only reduce \dg.


\subsection{Time tag}
It will be interesting if
very late decaying $B_s$ samples can be collected  \cite{Dun}.
Then, all the short
lived $B_s$ have already decayed and the beam
consists purely of the long lived state, $B_{long}$.
Of course, for any useful
measurement, large statistics is needed.
The expected number of reconstructed $B_s \to J/\psi \phi$ events is
about $1.2 \times 10^4$ at CDF in run II \cite{DeJongh},
and about $3.8 \times 10^5$ per year at the LHC \cite{Nev}.
Assuming that this can be achieved, $\xi$ could be measured
using the time tag method.

{}From Eqs. (\ref{slrate}) and (\ref{cprate}) we get \cite{Dun}
\begin{equation}
R(t) \equiv {\Gamma\left[f\left(t\right)\right]
\over \Gamma\left[g\left(t\right)\right]} \propto
\left\{ 1 - Re\,\lambda
\, \tanh \left(\frac{\Delta\Gamma\,t}{2}\right)\right\} \,,
\end{equation}
where $f$ is a CP even eigenstate and $g$ is a flavor specific state.
To enhance statistics, a sum over several
CP even final states can be performed.
Of course, if the time dependence can be traced,
\dg\ and $\xi$ can be extracted (recall, $Re\,\lambda=\cos 2\xi$).
In practice it might be easier just 
to make a cut to select events at large $t$, when 
$\tanh \left(\dg\, t/2\right) \to 1$.
This can be thought of as an almost pure $B_{long}$ beam.
Then $\xi$ can be obtained from the ratio
\beq
{R(t \to \infty) \over R(t \to 0)}=
{\Gamma(B_{long} \to f) \over 
\Gamma(B_s \to f)} = \left(1-|\cos 2\xi|\right)\,.
\eeq

Finally, let us estimate the loss in statistics if this method
is used. We define
\beq
P(t) \equiv {N(B_{long},t) \over N(B_{short},t)+N(B_{long},t)}=
\Big(1+\exp({-\dg\, t})\Big)^{-1}.
\eeq
where $N(B_a,t) \propto \exp(-\Gamma_a\,t)$ 
is the number of $B_a$ in the beam ($a=long,\;short$).
In order to get a data sample with purity $P$ for
a given \dg\ we need to make the cut at $t_{cut}$ such that
\beq
\Gamma \,t_{cut}>\left({\Gamma\over\dg}\right)\log\left({P\over1-P}\right).
\eeq
The loss in statistics is $\exp(-\Gamma \,t_{cut})$. For example,
for $P=92\%$ and $\dg/\Gamma = 0.30$ one has to wait
about 8 lifetimes, a loss of about $3 \times 10^3$ in statistics.
We learn that it may be possible to use the time tag method.


\subsection{Measuring $\gamma$}
In \cite{Dun} several  methods had been proposed for extracting 
\beq
\gamma \equiv\arg\left(-{V_{ud}V_{ub}^*\over
V_{cd}V_{cb}^*}\right).
\eeq
Here we briefly describe the methods and show how they are modified in the
presence of NP. More details on the methods, 
including the validity of the assumptions that
are used, are given in Ref. \cite{Dun}. 

There are basically two classes of methods.
The first one is based on decays into  CP eigenstates mediated by the
$b \to u \bar u d$ transition, e.g. $\rho^0 K_S$. 
When $\Gamma$ and \dg\ are known, 
a one parameter fit to the time dependence of untagged
$\rho^0 K_S$ events determines $Re \lambda$ (see Eq. (\ref{cprate})).
Assuming that the penguin contamination is small,
\beq
\lambda= \frac{q}{p}\,\frac{ \langle \rho^0 K_S\mid\overline{B}_s
\rangle }{ \langle \rho^0 K_S\mid B_s \rangle}= 
-e^{2i(\xi+\gamma)}.
\eeq
Then, $Re \lambda=-\cos(2\xi+2\gamma)$ and $\xi+\gamma$ can be extracted.

The second class uses pairs of final states
that are CP conjugate of each other and are
mediated by the
$b \to c \bar u s$ and $b \to u \bar c s$ transitions,
e.g. $D_s^- K^+$ and $D_s^+ K^-$. 
When the total decay width, 
${\Gamma (B_s \rightarrow f)}$ is known \cite{Dun},
a three parameter fit to the time dependences of the decays into 
$f$ and $\overline f$ determines $|\lambda|$, $Re \lambda$ and 
$Re \overline{\lambda}$ (see Eqs. (\ref{genrate})).
In general,
\begin{equation} 
\lambda = \frac{q}{p}\,\frac{ \langle D_s^- K^+\mid\overline{B}_s
\rangle }{ \langle D_s^- K^+\mid B_s \rangle}=
|\lambda|\, e^{i(\delta+2\xi+\gamma)}, \qquad
\overline{\lambda} = \frac{p}{q} \,\frac{ \langle D_s^+ K^- \mid
B_s \rangle}{ \langle D_s^+ K^- \mid\overline{B}_s  \rangle }=
|\lambda|\, e^{i(\delta-2\xi-\gamma)},
\end{equation}
where $\delta$ is the strong phase. 
Then, $Re \lambda=|\lambda|\cos(\delta+2\xi+\gamma)$,
$Re \overline{\lambda}=|\lambda|\cos(\delta-2\xi-\gamma)$
and $2\xi+\gamma$ can be extracted.

We learn that 
when $\Gamma$, \dg\ and $\xi$ are known (from the 
combination of the flavor
specific and the $b \to c \bar c s$ CP eigenstates decays data)
all these methods can be used to measure $\gamma$ up to discrete ambiguities.


\section{Conclusions}
Various extensions of the Standard Model predict new significant
contributions to the $B_s$ mass difference. 
In general, such models introduce new sources of CP violation
beyond the single phase of the CKM matrix.
Still, in most cases, the $B_s$ decay rates are described by the SM. 
Nevertheless, the width difference can be significantly reduced.
Actually, such a reduction is an indication of CP violation:
the large SM prediction for \dg\ is based on the fact that
the decay width into CP even final states is larger than into CP odd 
final states.
When new CP violating phases appear in the mixing amplitude then
the mass eigenstates can sizably differ from the CP eigenstates, and
both mass eigenstates 
are allowed to decay into CP even final states\footnote{%
When $|q/p|=1$ one can
always choose a convention for CP transformation such that 
the mass eigenstates are CP eigenstates. But then,
CP is violated by the decay amplitudes and both CP eigenstates
decay into CP even states. Of course,
the final result is the same, no matter what convention  we use.}.
Consequently, \dg\ reduced.

Since in the SM the CP violating phase $\xi$ is very small \cite{Yossi},
($\xi\le|V_{us}V_{ub}/V_{cs} V_{cb}|<2.5\times10^{-2}$),
a measurement of $\xi \ne 0$ would be a clear evidence for 
CP violation beyond the SM. 
Actually, $\xi \ne 0$ is the
equivalent of $\alpha+\beta+\gamma \ne \pi$, which
seems to be much harder to test experimentally \cite {nisi}.
Furthermore, $\xi$ can be extracted using the $B_s$ 
leading $b \to c \bar c s$ decay modes (CKM unsuppressed).
All these reasons explain why the measurement of $\xi$ is very important.
If the rapid $\dm t$ oscillation can be
traced, the time dependent CP asymmetries 
in $B_s$ decay modes mediated by $b \to c \bar c s$ 
will measure $\sin 2\xi$. 
However, untagged $B_s$ samples seem to be the best place 
for this measurement.
We described several ways of determine $\xi$
using untagged $B_s$ samples: 
\begin{itemize}
\item{%
Compare the SM calculation (which has large
uncertainties) to the measured \dg,
\beq
|\cos 2\xi| ={\dg \over \dg_{SM}}.
\eeq
}
\item{%
Compare \dg\ as measured from decays into CP eigenstates,
to \dg\ from flavor specific decays,
\beq
|\cos 2\xi| = {\dg_{CP} \over \dg_{FS}}.
\eeq
}
\item{%
If a measurement at late times is possible, decay rates of
a specific mass eigenstate can be measured, and then
\beq
|\cos 2\xi| = 1-{\Gamma(B_{long} \to {\rm CP\ even}) \over \Gamma(B_s \to
{\rm CP\ even})}.
\eeq
}
\end{itemize}


\acknowledgments
I thank Francesca Borzumati, Isi Dunietz,
Neville Harnew, Zvi Lipkin, Louis Lyons, Enrico Nardi and Yossi Nir 
for helpful discussions and comments.


{\tighten

}  

\end{document}